\documentclass[aps,prl,reprint,showpacs,preprintnumbers,amsmath,amssymb,citeautoscript]{revtex4-1} 

\usepackage{graphicx}
\graphicspath{{figs/}{./}}

\usepackage{color}
\newcommand\FigureFile[1] {#1.eps}
\DeclareGraphicsExtensions{pdf}

\usepackage{dcolumn}
\usepackage{bm}
\usepackage{color}
\usepackage{multirow}

\newcommand\eq[1]                              
{
\begin{align*}
#1
\end{align*}
}

\newcommand\eql[2] 
{
\begin{equation}\label{#1}
\begin{split}
#2
\end{split}
\end{equation}
}

\newcommand\eqsl[1]                            
{
\begin{align}
#1
\end{align}
}

\newcommand\eqssl[2]                      
{
\begin{subequations}\label{#1}
\begin{align}
#2
\end{align}
\end{subequations}
}

\definecolor{xmgrace-green4}{rgb}{0.0,0.55,0.0}
\definecolor{Green}{rgb}{0.2,0.96,0.2}
\definecolor{Remarks}{rgb}{1,0.3,0.3}
\definecolor{Extra}{rgb}{0.2,0.2,1}
\definecolor{Blue}{rgb}{0.2,0.3,1}
\definecolor{Black}{rgb}{0,0,0}

\newcommand\compPackage[1] {{\footnotesize{#1}}}
\newcommand\NWCHEM     {\compPackage{NWCHEM}}
\newcommand\QUANTUMESPRESSO   {\compPackage{QUANTUM ESPRESSO}}

\newcommand\PWSCF      {\compPackage{PWSCF}}

\newcommand\COMMENTED[1] {}

\begin{document}

\title{The Stability, Energetics, and Magnetic States of Cobalt Adatoms on Graphene}

\author{Yudistira Virgus}
\email{yvirgus@email.wm.edu}
\affiliation{Department of Physics, College of William and Mary,
Williamsburg, Virginia 23187-8795, USA}

\author{Wirawan Purwanto}
\affiliation{Department of Physics, College of William and Mary,
Williamsburg, Virginia 23187-8795, USA}

\author{Henry Krakauer}
\affiliation{Department of Physics, College of William and Mary,
Williamsburg, Virginia 23187-8795, USA}

\author{Shiwei Zhang}
\affiliation{Department of Physics, College of William and Mary,
Williamsburg, Virginia 23187-8795, USA}

\date{July 2, 2014}

\begin{abstract}

We investigate the stability and electronic properties of single Co
atoms on graphene with near-exact many-body calculations. A
frozen-orbital embedding scheme was combined with auxiliary-field quantum
Monte Carlo to increase the reach in system sizes.  Several energy minima are
found as a function of the distance $h$ between Co and graphene.  
Energetics only permit the Co atom to occupy the top site at $h = 2.2$~\AA\ 
in a  high-spin $3d^{8}4s^{1}$ state, and
the van der Waals region at $h = 3.3$~\AA\ in a high-spin $3d^{7}4s^{2}$ state.  
The findings provide an explanation for recent experimental results with Co on free-standing graphene.
 \end{abstract}


\maketitle

Graphene, with its unique band structure at the Dirac point and exceptional physical properties, has the potential to revolutionize electronics technology \cite{Geim2007, Castro2009, Novoselov2012}.
Recently, research interests in the adsorption of transition metal adatoms on graphene 
have grown rapidly because of its possible use to induce magnetism for spintronic applications \cite{Pesin_2012, Yazyev_2010}.
Single Co atoms on graphene have been extensively studied, both theoretically \cite{Yagi2004, Mao2008, Johll2009, Wehling2010-a, Wehling2010-b, Jacob2010, Cao2010, Valencia2010, Chan2011, Liu2011, Sargolzaei2011, Ding2011, Wehling2011,Rudenko2012,Virgus2012} and experimentally \cite{Brar2011,Eelbo2013a, Eelbo2013b,Donati2013,Sessi2014}.
For example, scanning tunneling microscopy (STM) experiments have demonstrated the ability to controllably ionize a Co adatom on graphene using a back gate voltage \cite{Brar2011}.
A high magnetic anisotropy for Co/graphene has been observed \cite{Donati2013}.  
It is thus of great importance to understand the properties of Co/graphene both from a fundamental and applied perspective.

Most theoretical studies have addressed Co adsorption on graphene at the density functional theory (DFT) level, using local or semi-local functionals, or an empirical Hubbard on-site repulsion $U$ (DFT$+U$)  \cite{Yagi2004, Mao2008, Johll2009, Wehling2010-a, Wehling2010-b, Jacob2010, Cao2010, Valencia2010, Chan2011, Liu2011, Sargolzaei2011, Ding2011, Wehling2011}. 
Although these approaches have often given reasonable results in a variety of materials, 
indications are that in the system of interest, combining a transition metal and graphene, their accuracy is uncertain.
Indeed calculations have reported qualititatively different results for the nature of the magnetic state, adsorption site, and binding energy of Co as a function of adsorption height.
For example, DFT using the generalized gradient approximation (GGA) \cite{Perdew1996} shows \cite{Yagi2004, Mao2008, Johll2009, Wehling2010-a, Wehling2010-b, Jacob2010, Cao2010, Valencia2010, Chan2011, Liu2011, Sargolzaei2011, Ding2011, Wehling2011} that the sixfold hollow site is the global minimum, with an equilibrium height of $h_{\textrm{eq}} \sim 1.5$~\AA\ and a low-spin  $3d^{9}4s^{0}$ Co atom configuration ($S=1/2$).
A different functional, the hybrid Becke three-parameter Lee-Yang-Parr  (B3LYP) \cite{Becke1993}, predicts \cite{Jacob2010} an equilibrium height of $h_{\textrm{eq}} \sim 1.9$~\AA\ with a high-spin  $3d^{8}4s^{1}$ configuration ($S=3/2$) at the hollow site. 
The GGA+$U$ approximation has shown sensitivity to the value of the parameter $U$ used. For $U=2$ eV, it predicts the global minimum to be the hollow site with a low-spin configuration, while for $U=4$ eV, the global minimum is the top site with a high-spin configuration \cite{Wehling2010-b, Chan2011, Wehling2011}.
A quantum chemistry calculation using the complete active space self-consistent field (CASSCF) method predicts the global minimum at the van der Waals (vdW) region with $h_{\textrm{eq}} \sim 3.1$ \AA\ and a high-spin $3d^{7}4s^{2}$ configuration \cite{Rudenko2012}.
The varying results underscore the need for better understanding of and fundamentally more accurate 
 approaches to treat transition metal adsorption on graphene.

In this paper we address the problem from two complementary angles, using auxiliary-field quantum Monte Carlo (AFQMC)  \cite{Zhang1997_CPMC, Zhang2003,Zhang2013} calculations. 
First we apply an exact free-projection AFQMC  approach to systematically benchmark the various theoretical methods  in a series of model systems which are smaller in size but retain key features of Co/graphene.
These results will provide guidance for future studies of transition metal adsorption on graphene, especially in the selection of computationally less costly approaches.
Secondly, a frozen-orbital embedding scheme is developed to extend the system size that can be treated with AFQMC.
Using the new approach, we determine  the property for Co/graphene by
direct AFQMC calculations of Co on large substrates 
(e.g., C$_{24}$H$_{12}$, seven hexagonal carbon rings), augmented by a finite-size 
correction from the substrate to graphene, treated by DFT.
As discussed below, our results are consistent with and provide a quantitative explanation for the observations from recent STM experiments of Co adatoms adsorbed on H-intercalated graphene/SiC(0001) \cite{Eelbo2013b}. 

Most DFT calculations of Co/graphene give the hollow site as the global minimum.  
Using AFQMC, we had determined \cite{Virgus2012} that bonding at the hollow site, 
 different from the DFT predictions, exhibited a double-well structure with nearly equal binding energy.
Recently, however, experimental studies have indicated that single Co atoms  can be adsorbed on both the hollow site and the top site \cite{Eelbo2013a, Eelbo2013b, Donati2013}.
Motivated by these results, here we investigate the binding energy and electronic properties of Co/graphene  for all three high-symmetry adsorption sites: the sixfold hollow site, the twofold bridge site, and the top site. 
We find that, among the different energy minima with different electronic configurations and adsorption sites,
 only two are stable and can be occupied by the Co atom under experimental 
conditions.
The first minimum corresponds to the vdW interaction, while the other is at the top site which arises from strong orbital hybridization.  

The AFQMC method \cite{Zhang1997_CPMC, Zhang2003} stochastically evaluates the ground-state properties of a many-body Hamiltonian by means of random walks with Slater determinants expressed in a chosen single-particle basis. 	
While exact in principle, the fermion sign problem causes in exponential growth of the Monte Carlo variance. 
The problem is controlled using the phaseless approximation \cite{Zhang2003}, which imposes a constraint on the overall phase of the Slater determinants using a trial wave function  $\Psi_\mathrm{T}$.
Phaseless AFQMC has demonstrated excellent accuracy in a wide variety of molecular and crystalline systems and also strongly correlated lattice models  \cite{Zhang1997_CPMC, Zhang2003,AlSaidi2006a,AlSaidi2006b,AlSaidi2006c,AlSaidi2007b,Purwanto2008,Zhang2013},
often with simple forms of $\Psi_\mathrm{T}$.
The sign problem can also be attacked directly by lifting or releasing the constraint, using a large number of random walkers.
This approach is exact, although exponentially scaling in computational cost with system size, and will be 
referred to as  free-projection AFQMC (FP-AFQMC) \cite{Zhang2003,Purwanto2009_Si,Hao_Shi2013}. 

In this work we implement a frozen-orbital approach to allow direct 
AFQMC calculations on large system sizes, for example, Co on coronene (Co/C$_{24}$H$_{12}$). 
A size-correction embedding scheme is then employed to remove the residual difference between  Co/coronene and Co/graphene. 
 Because strong electron-electron correlation effects are  spatially localized in the vicinity of the Co atom, 
it is sufficient to treat the  size correction with a lower level of theory. We have used DFT for the 
residual finite-size corrections, which are found to be very small, as further discussed below.

In the frozen-orbital AFQMC approach, the molecular orbitals of the cluster, e.g., Co/C$_{24}$H$_{12}$, which are obtained from Hartree-Fock (HF), are transformed into localized orbitals. 
The Foster-Boys method \cite{Boys1960} is employed for the orbital localization, as implemented in  {\NWCHEM} \cite{Valiev2010}.
In the AFQMC calculations, the C-H bonds and the outer-most C-C bonds are then frozen, using a formalism similar to frozen-core \cite{Purwanto2013}. 
This accelerates the many-body calculations greatly while introducing essentially no error in the binding energy. 

\begin{table}[tbp]
\caption{\label{tbl:Co-C6H6}
Calculated binding energies of Co on C$_6$H$_6$ at three heights near the local minima. 
The uncertainty on AFQMC results includes both statistical and systematic errors and the uncertainty for CCSD(T) results comes from CBS extrapolations.
}
\begin{ruledtabular}
\begin{tabular}{c c c c}
\vspace{-1.1em}
\\
                       & \multicolumn{3}{c}{Binding Energy (eV)}\\ \cline{2-4}
\vspace{-0.8em}
\\
                        & FP-AFQMC & AFQMC &   CCSD(T) \\
\hline
\vspace{-0.8em}
\\
$S=1/2$ ($h=1.5$ \AA)\
                      & $ -0.99(4)$ &   $-1.07(4)$  & $-0.91(4)$              \\
\hline
\vspace{-0.8em}
\\
$S=3/2$ \\

$3d^{8}4s^{1}$ ($h=1.7$ \AA)\              
                    &  $-0.92(4)$ & $-0.91(4)$  & $-0.95(4)$\\
vdW ($h=3.0$ \AA)\
                     &  $-0.15(3)$   & $-0.15(3)$  & $-0.15(3)$\\
\end{tabular}
\end{ruledtabular}
\end{table}

\begin{figure}[tbp]
\includegraphics[scale=0.93]{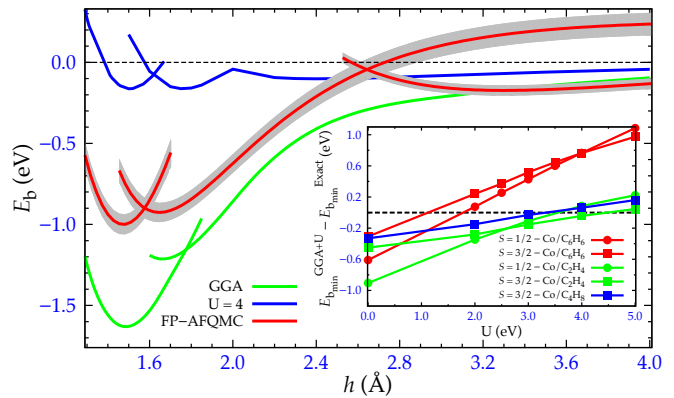}
\caption{\label{fig:Co-C6H6-AFQMC-GGA-U-y0-inset-3models}
(Color online) Binding energy of Co on C$_6$H$_6$ as a function of Co adsorption height $h$ at the six-fold site for GGA, GGA+$U$, and exact FP-AFQMC. 
For FP-AFQMC, left, middle, and right curves correspond to nominal $3d^{9}4s^{0}$, $3d^{8}4s^{1}$, and $3d^{7}4s^{2}$ Co configurations, respectively. 
For GGA and GGA+$U$, the left and right curves correspond to $3d^{9}4s^{0}$ and $3d^{8}4s^{1}$ Co configurations, respectively. 
The shaded area on the AFQMC Morse fits reflects one standard deviation which includes both statistical and systematic errors.
The inset shows GGA+$U$ errors as a function of $U$.
The difference in the calculated binding energy between GGA+$U$ and exact results 
is plotted for Co/C$_6$H$_6$, Co/C$_2$H$_4$, and Co/C$_4$H$_8$, 
at their respective minimum position $h$.   
}
\end{figure}

\begin{figure}[tbp]
\includegraphics[scale=0.79]{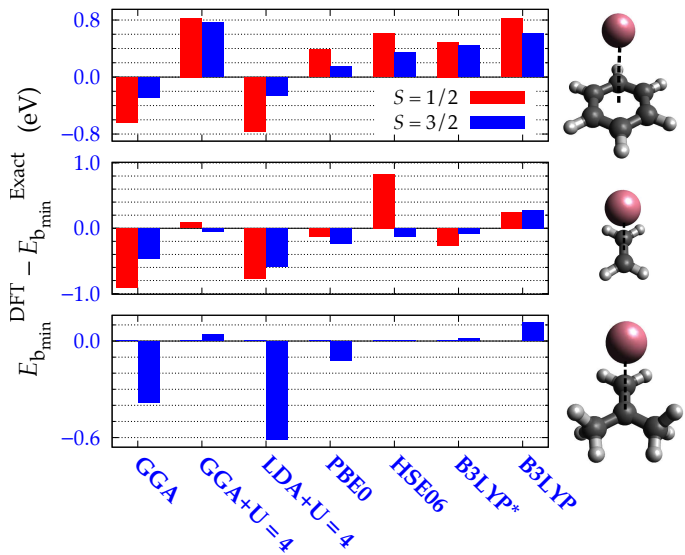}
\caption{\label{fig:Co-C6H6-Co-C2H4-Co-C4H8-models}
(Color online) The error in the calculated binding energy from DFT, for Co/C$_6$H$_6$, Co/C$_2$H$_4$, and Co/C$_4$H$_8$, respectively.
The low-spin ($S=1/2$) and high-spin ($S=3/2$) states correspond to $3d^{9}4s^{0}$ and $3d^{8}4s^{1}$ Co configurations, respectively. 
The results from GGA+$U$ and LDA+$U$ for $U=4$ eV are also shown for comparison.
}
\end{figure}

We first use FP-AFQMC to obtain exact results in three model systems: Co/C$_6$H$_6$, Co/C$_2$H$_4$, and Co/C$_4$H$_8$, which represent prototypes of the three high-symmetry adsorption sites in Co/graphene. 
These results are used to benchmark the phaseless AFQMC,  coupled-cluster \cite{Bartlett2007} [CCSD(T)], and various DFT methods.
The C-C bond length in the model systems was fixed to that of graphene,  $1.42$ \AA,\ and the C-H bond length was set to $1.09$  \AA,\  which is close to the experimental bond lengths in the three molecules. 
The results are summarized in Table~\ref{tbl:Co-C6H6}, 
Fig.~\ref{fig:Co-C6H6-AFQMC-GGA-U-y0-inset-3models}, and Fig.~\ref{fig:Co-C6H6-Co-C2H4-Co-C4H8-models}. 
In the calculations, we employ Gaussian basis sets for DFT, hybrid DFT, and HF, which were performed with {\NWCHEM}. 
DFT+$U$ calculations were done with
the {\PWSCF} code of the {\QUANTUMESPRESSO} package \cite{Giannozzi2009}, using planewaves and  ultrasoft pseudopotentials \footnote{We obtained the pseudopotentials from http://www.quantum-espresso.org; H.pbe-rrkjus.UPF, C.pbe-rrkjus.UPF, and Co.pbe-nd-rrkjus.UPF for GGA calculations and H.pz-rrkjus\_psl.0.UPF, C.pz-n-rrkjus\_psl.0.UPF, and Co.pz-nd-rrkjus.UPF for LDA's},
with a $50$\,Ry kinetic energy cutoff and a charge density cutoff of 400\,Ry.
Each of the three model systems, the substrate molecules, and the Co atom were treated using a
15\,\AA\ cubic supercell. 
The AFQMC calculations also used standard Gaussian basis sets and a frozen-core approximation to treat the inner core electrons \cite{Purwanto2013}.
The basis sets and other run parameters and the procedure for extrapolation to the complete basis set (CBS) limit were similar to those in Ref. \onlinecite{Virgus2012}. 
A single-determinant HF trial wave function (or its equivalent after a localization transformation of the 
occupied orbitals in the frozen-orbital calculations) was used in most cases for high-spin states, while a multideterminant  $\Psi_\mathrm{T}$ obtained from CASSCF was often applied 
in states involving low-spin Co atom configuration.

We find that phaseless AFQMC and CCSD(T) produce accurate binding energy curves for high-spin Co atom configurations ($S = 3/2$) in all three model systems. 
In the case of the low-spin configuration ($S = 1/2$), both show small errors,
with phaseless AFQMC overestimating the binding energy of Co/C$_6$H$_6$ by $\sim 0.1$\,eV while CCSD(T) underestimating it by approximately the same amount, as shown in Table~\ref{tbl:Co-C6H6}.

We also benchmark DFT, with local and hybrid functionals, and DFT$+U$ methods.
Figure~\ref{fig:Co-C6H6-AFQMC-GGA-U-y0-inset-3models} shows the binding energy curves of Co/C$_6$H$_6$ as a function of $h$ from GGA, GGA+$U$, in comparison with exact  FP-AFQMC results. 
All energies have been extrapolated to the CBS limit. 
The FP-AFQMC results show that the ground-state electronic configuration of the Co atom undergoes two transitions as $h$ decreases, which produces three different configurations: high-spin $3d^{7}4s^{2}$,  high-spin $3d^{8}4s^{1}$, and low-spin $3d^{9}4s^{0}$ states, respectively.
All DFT functionals and DFT+$U$ produce only two ground-state configurations, a  high-spin $3d^{8}4s^{1}$ for high $h$'s and a low-spin $3d^{9}4s^{0}$ for small $h$'s, since they incorrectly predict the $3d^{8}4s^{1}$ state as the ground-state configuration for the free Co atom.

Figure~\ref{fig:Co-C6H6-Co-C2H4-Co-C4H8-models}  summarizes the error in the calculated binding energy (near the optimal geometry as determined 
by FP-AFQMC)
from DFT using different functionals for all three model systems.
In Co/C$_2$H$_4$, the same three nominal ground-state electronic configurations of the Co atom are found as those in Co/C$_6$H$_6$.
In Co/C$_4$H$_8$, however,  the low-spin $3d^{9}4s^{0}$ state is unbound, so only the high-spin states are considered.
We chose a representative set of the most common DFT functionals \cite{Kieron2012}: the local density approximation (LDA), GGA, the hybrid Perdew-Burke-Ernzerhof (PBE0), the hybrid Heyd-Scuseria-Ernzerhof (HSE06), B3LYP, and a modification for transition metals, B3LYP*.
As seen from the figure, none of the functionals gives a uniformly correct description of all three model systems.
The hybrid PBE0 and B3LYP* show the best agreement among the DFT functionals.
 
Two typical DFT+$U$ results are also included for comparison in Fig.~\ref{fig:Co-C6H6-Co-C2H4-Co-C4H8-models}.  
A more systematic analysis of the accuracy of  DFT+$U$ is given in the inset in Fig.~\ref{fig:Co-C6H6-AFQMC-GGA-U-y0-inset-3models}, in
which we test the range of values for the parameter $U$ determined from Refs. \onlinecite{Wehling2010-b, Chan2011, Wehling2011} (with $J=0.9$ eV).
The results suggest that there is no single ``correct'' value of $U$ that can quantitatively capture 
the physics of
Co adsorption on graphene across the different configurations.  
For Co/C$_6$H$_6$, $U \sim 2$ eV shows good agreement with the exact result, while Co/C$_2$H$_4$ and Co/C$_4$H$_8$ require larger values, $U \sim 4$ eV.

We next determine the properties of Co/graphene using frozen-orbital phaseless AFQMC. 
Our tests show that, to reach the desired accuracy in predicting the binding energies in Co/graphene, 
 the model systems above are inadequate to use as the near-regions for an 
embedding treatment. 
Instead we use Co/C$_{24}$H$_{12}$, Co/C$_{10}$H$_{8}$ (two hexagonal carbon rings), and Co/C$_{13}$H$_{10}$ (three hexagonal carbon rings) as the near regions for the hollow, the bridge, and the top sites, respectively. 
In the frozen-orbital AFQMC calculations, 
we freeze the bonds farthest from the Co atom. 
To validate the results, select AFQMC calculations are carried out on the entire cluster, 
for certain basis sets and geometries, to compare with the corresponding frozen-orbital results. 
The calculated binding energies agree within statistical error bars.

The residual finite-size corrections from the near-region clusters to Co/graphene 
are treated with DFT. 
The final binding energy, after size-correction, of Co/graphene at each geometry is given by
\eql{eq:ONIOM}
{
 E_{\textrm{\scriptsize{{b}}}}^{\textrm{\scriptsize{Co/graphene}}} = E_{\textrm{\scriptsize{{b}, Exact}}}^{\scriptsize{\textrm{Co/}z}} + ( E_{\textrm{\scriptsize{{b}, DFT}}}^{\textrm{\scriptsize{Co/graphene}}} -  E_{\textrm{\scriptsize{{b}, DFT}}}^{\scriptsize{\textrm{Co/}z}})       
       \,,
}
where $z$ denotes the near region's substrate and geometry. 
The binding energies are defined as $E_{\textrm{\scriptsize{{b}}}}(h) \equiv E^{\textrm{\scriptsize{Co/substrate}}}(h) -  E^{\textrm{\scriptsize{Co}}} - E^{\textrm{\scriptsize{substrate}}} $, where the last two terms on the right are the total energies of the isolated Co atom and corresponding
substrate, respectively.
The Co/graphene DFT binding energy  was obtained from {\PWSCF} calculations using a \mbox{5 $\times$ 5} in-plane supercell, which contains 50 C atoms and a Co atom.  
The in-plane lattice parameter was $12.3$ \AA\  and the out-of-plane distance perpendicular to the graphene plane was set to $15$ \AA.
Brillouin-zone sampling used a $\Gamma$-centered  \mbox{$ 4 \times 4 \times 1$}  $k$-point grid and a Gaussian smearing of $0.04$ eV.
Planewave cutoffs were as in the DFT+$U$ calculations above.
Substrate relaxation effects were included for each $h$ as an additional 
size-correction layer, and relaxation was considered complete when the forces on all atoms were less than $0.02$ eV/\AA.
We have checked that the size-correction  in Eq.~(\ref{eq:ONIOM}) is insensitive to the choice of DFT exchange-correlation functional, even though the different functionals differ in their description of the components \cite{Virgus2012}.

\begin{figure}[tbp]
\includegraphics[scale=0.95]{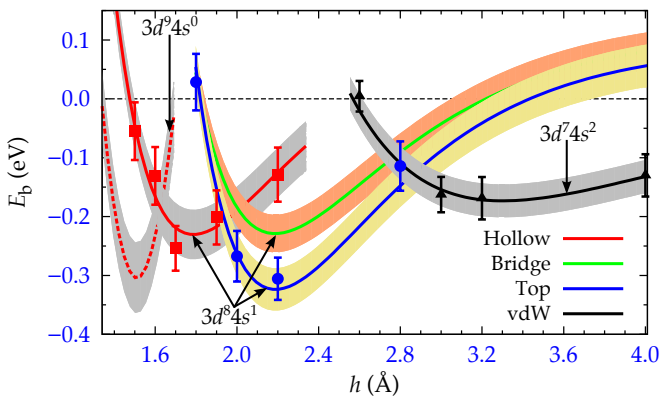}
\caption{\label{fig:Co-graphene-oniom-CBS-relaxed}
(Color online) Binding energy of Co on graphene as a function of $h$ for the three sites.
The left, middle, and right curves correspond to $3d^{9}4s^{0}$, $3d^{8}4s^{1}$, and $3d^{7}4s^{2}$ Co configurations, respectively. 
The dashed line indicates the low-spin state Co atom.
Extrapolation to the CBS limit has been included.
Shaded areas are one-$\sigma$ estimates of uncertainties, including the statistical errors in AFQMC.
}
\end{figure}

The final binding energy curves of Co/graphene for all three sites are shown in Fig.~\ref{fig:Co-graphene-oniom-CBS-relaxed}.
All results are obtained with AFQMC except for the bridge site which is provided by CCSD(T).
Several CCSD(T) calculations are also done for the vdW region and the top site and they agree with AFQMC results, consistent with the benchmark results earlier.
The lines in Fig.~\ref{fig:Co-graphene-oniom-CBS-relaxed} are Morse fits to the AFQMC and CCSD(T) results.
The size correction is not applied to the vdW region because the chosen DFT functionals did not include
 vdW interactions (see Ref.~\onlinecite{vdW-Berland2014}, for example).
(Indeed, with the standard functionals used here, the vdW region becomes unbound after size-correction is applied \cite{Virgus2012}.) 
The vdW binding curve in Fig.~\ref{fig:Co-graphene-oniom-CBS-relaxed}  is expected to be free of finite-size errors and nearly exact, however,
since both AFQMC and CCSD(T) calculations show Co/C$_6$H$_6$ to be close to Co/C$_{24}$H$_{12}$, Co/C$_{10}$H$_{8}$, and Co/C$_{13}$H$_{10}$, which give essentially the same binding energy.
For reference, we also show the low-spin $3d^{9}4s^{0}$ hollow-site curve (dashed line) obtained from earlier phaseless AFQMC calculations \cite{Virgus2012}, after correcting the phaseless bias with the FP results of Table~\ref{tbl:Co-C6H6}.
Because of energy barriers, this curve will not be relevant to the energetics, as discussed below.

Multiple energy minima are seen in Fig.~\ref{fig:Co-graphene-oniom-CBS-relaxed}, associated with different adsorption sites and different spin states.
However, a closer examination of the energetics shows that the Co atom can only occupy 
the vdW region (all sites) at $h \sim 3.3$ \AA\ and 
the top site at $h \sim 2.2$ \AA.
The binding energy at the vdW region is $\sim -0.18$\,eV, while that at the 
 top site is  $\sim -0.31$\,eV. 
Although the bridge site and hollow sites are also bound in the nominal $3d^{8}4s^{1}$ configuration, they 
are either metastable or inaccessible in the adsorption process.
Test calculations for Co/C$_{13}$H$_{10}$ indicate that the bridge site is a saddle point, unstable to Co relaxing to the top site.
The minimum for the hollow site lies at smaller $h \sim 1.7$\,\AA.
The kinetic barrier between it and the top site is $\sim 0.13$ eV, which is very large compared with the temperatures at which the experiments are performed \cite{Brar2011,Eelbo2013a, Eelbo2013b,Donati2013,Sessi2014}. 
Energetically, the Co atom is prevented, therefore, from hopping to the inner hollow site.
(In contrast, the barrier height between the vdW region and the top site  is much smaller.) 

A recent STM experiment for Co atoms on H-intercalated graphene/SiC(0001), also called quasi-free-standing monolayer graphene (QFMLG), reported that single Co atoms can be adsorbed at the top site and the hollow site with $h=2.2$ \AA\  and  $h=3.1$ \AA,\ respectively \cite{Eelbo2013b}.
The finding is in close agreement with our results.  
The vdW region in our calculations can be associated with the hollow site, since any Co atoms in the vdW region at the top and bridge sites can easily hop to the top site global minimum due to low kinetic barriers, whereas Co atoms in the vdW hollow site cannot hop to the inner hollow site.
The experiment also showed that Co atoms at the hollow site switched to the top site at a bias voltage of $\sim -0.25$ V, which is of the order of the vdW binding energy. This could allow Co atoms trapped at the hollow site vdW minimum to migrate to the top site.

Experimental studies of  Co on graphene/SiC(0001), dubbed monolayer graphene (MLG),
observed only the top site \cite{Eelbo2013a,Eelbo2013b}, however. 
This difference might arise from the fact that MLG is less well modeled by the free graphene system 
studied here. MLG, grown on top of a carbon buffer layer, is strongly \emph{n}-doped due to the interaction with the buffer layer, which can cause more deviations from the linear dispersion near the Dirac point \cite{Qi2010,Forti2011} than in QFMLG (which is only 
 slightly \emph{p}-doped   \cite{Riedl2009,Forti2011}).
Furthermore, MLG shows significant corrugations, while QFMLG is exceptionally flat \cite{Forti2011,Goler2013}. 
Similar considerations may apply to STM experiments with Co on graphene/Pt(111), which observed only the hollow site at $h = 2.4$ \AA \ \cite{Donati2013}.
Although the interaction of graphene and Pt(111) is assumed to be weak \cite{Preobrajenski2008,Sutter2009}, experimental studies suggest that there likely exists hybridization between graphene Dirac cone states and Pt \emph{d} orbitals \cite{Rajasekaran2012,Zhou2013}. 
Further investigations, both experimental and theoretical, are needed to resolve these issues and the substrate effects on graphene.

In summary, we have presented highly accurate many-body results on the adsorption of Co on graphene for the three high-symmetry sites.
With model systems Co/C$_6$H$_6$, Co/C$_2$H$_4$, and Co/C$_4$H$_8$, exact results are obtained. 
Our benchmark study showed that phaseless AFQMC and CCSD(T) are essentially exact for high-spin Co configurations. 
DFT with various functionals and DFT+$U$ give widely varying results, 
cautioning that care must be taken in future studies of transition metal  on graphene using
such approaches. A quantitative measure of the accuracy is provided for the most commonly used functionals
and for the choice of $U$ values.
A frozen-orbital AFQMC approach was introduced to allow direct many-body calculations on 
large clusters, Co/C$_{24}$H$_{12}$, Co/C$_{10}$H$_{8}$, and Co/C$_{13}$H$_{10}$. 
A size-correction embedding scheme was then employed to calculate the binding energy of Co/graphene.
We find that the Co atom can be adsorbed at the top site with  \mbox{$E_\mathrm{b} \sim -0.31$\,eV} and at the vdW region with  \mbox{$E_\mathrm{b} \sim -0.18$\,eV}.
The results explain recent experimental observations for Co on H-intercalated graphene/SiC.

This work was supported by 
DOE (DE-FG02-09ER16046), 
NSF (DMR-1409510),
and 
ONR (N000140811235; N000141211042).
An award of computer time was provided by the Innovative and Novel Computational Impact on Theory and Experiment (INCITE) program, using resources of the Oak Ridge Leadership Computing Facility at the Oak Ridge National Laboratory, which is supported by the Office of Science of the U.S. Department of Energy under Contract No. DE-AC05-00OR22725.
We also acknowledge computing support from the Blue Waters at UIUC supported by NSF PRAC, and the SciClone Cluster at the College of William and Mary.

\bibliography{AFQMC-bib-entries,Co-graphene,Miscellaneous}

\end{document}